\documentclass[aps,prb,twocolumn,superscriptaddress,longbibliography,nobibnotes,nodoi,nourl,noeprint]{revtex4-2}
\usepackage{amsmath,amssymb,physics,bm,siunitx}
\usepackage{xcolor,graphicx,soul}
\usepackage[colorlinks=true,citecolor=blue,urlcolor=blue,linkcolor=black]{hyperref}

\newcommand{\kb}{k_{\scriptscriptstyle\rm B}}

\def\rms{\rm\scriptscriptstyle} \def\dd{{\rm d}}

\definecolor{darkred}{rgb}{0.7, 0, 0}

\begin{document}
	
\title{Universal time-temperature scaling of conductivities in random site energy\\ and associated random barrier model}

\author{Sven Lohmann}
\email{slohmann@uni-osnabrueck.de}
\affiliation{Universit\"{a}t Osnabr\"uck, Institut f\"ur Physik, Barbarastra{\ss}e 7, D-49076 Osnabr\"uck, Germany}

\author{Quinn Emilia Fischer}
\email{qfischer@uni-osnabrueck.de}
\affiliation{Universit\"{a}t Osnabr\"uck, Institut f\"ur Physik, Barbarastra{\ss}e 7, D-49076 Osnabr\"uck, Germany}

\author{Justus Leiber}
\email{jleiber@uni-osnabrueck.de}
\affiliation{Universit\"{a}t Osnabr\"uck, Institut f\"ur Physik, Barbarastra{\ss}e 7, D-49076 Osnabr\"uck, Germany}

\author{Philipp Maass} 
\email{maass@uni-osnabrueck.de}
\affiliation{Universit\"{a}t Osnabr\"uck, Institut f\"ur Physik, Barbarastra{\ss}e 7, D-49076 Osnabr\"uck, Germany}

\date{January 13, 2026}

\begin{abstract}
Universal time-temperature scaling of conductivity spectra in
disordered solids has been explained by thermally activated hopping of
noninteracting particles over random energy barriers.  An open problem
is whether the random barrier model accounts for site energy disorder
in real materials.  Through mapping many-particle hopping in a
disordered site energy landscape to that of independent particles in a
barrier landscape, we show that time-temperature scaling is correctly
described by the associated barrier model in the low temperature
limit.  However, the site energy model displays good scaling behavior
at substantially higher temperatures than the barrier model, in
agreement with experimental observations.  Extending the mapping to
different types of mobile charge carriers allows us to understand why
time-temperature superposition can be absent in mixed alkali glasses.
\end{abstract}

\maketitle

\section{Introduction}
\label{sec:introduction}
Conductivities of disordered solid materials exhibits universal
scaling properties at frequencies below 100~GHz
\cite{Summerfield:1985, Dyre/Schroeder:2000}. Spectra
$\sigma(\omega)=\sigma'(\omega)+i\sigma''(\omega)$ at different
temperatures $T$ fall onto a single master curve when
$\sigma(\omega,T)$ is normalized to the dc conductivity $\sigma_{\rm
  dc}$ and the frequency scaled with $\sigma_{\rm dc}/\Delta\epsilon$:
\begin{equation}
\frac{\sigma(\omega,T)}{\sigma_{\rm dc}}
=F\left(\frac{\Delta\epsilon\,\omega}{\sigma_{\rm dc}}\right)\,.
\label{eq:sigma-BNN-scaling}
\end{equation}
Here, $\Delta\epsilon=\lim_{\omega\to0}[\sigma''(\omega,T)/\omega]$ is
the difference between the dielectric permittivities at low and high
frequencies.  Frequency-temperature scaling, also referred to as
time-temperature superposition principle, is observed for electron
conduction in amorphous semiconductors \cite{Summerfield:1985,
  Abboudy/etal:1988, BenChorin/etal:1995, Cil/Aktas:1991,
  Godet/etal:2007}, polymers \cite{Rehwald/etal:1987,
  Jastrzebska/etal:1998, Brom/etal:1998}, organic-inorganic hybrid
compounds \cite{Garoui/etal:2025}, and polymer-carbon nanotube thin
films \cite{Kilbride/etal:2002}, for polaron-mediated electron
transport \cite{Zuppiroli/etal:1991, Nagao/etal:2007,
  Ghodhbani/etal:2024}, as well as for ion conduction in polymer
electrolytes \cite{Rozanski/etal:1995, Khiar/etal:2006,
  Imre/etal:2009, Pal/Ghosh:2015} and glasses \cite{Roling/etal:1997,
  Roling:1998, Pan/Ghosh:1999, Roling/etal:1999, Belin/etal:2000,
  Ghosh/Pan:2000, Cramer/etal:2002, Cramer/etal:2003a,
  Dutta/Ghosh:2005, Bhattacharya/Ghosh:2007, Staesche/Roling:2010,
  Ali/Shaaban:2011, Deb/Ghosh:2011, Sangoro/Kremer:2012,
  Singh/etal:2012, Purnima/etal:2013, Christensen/etal:2013,
  Sklepic/etal2014, Chaurasia/etal:2015, Tripathy/etal:2015,
  Nasri/etal:2016, Singh/etal:2016, Gandi/etal:2019, Palui/Ghosh:2019,
  Sadok/etal:2019, Ojha/etal:2021, Ghosh/etal:2022, Makani/etal:2022,
  Mandal/etal:2023}.

The dc conductivity is generally thermally activated, $T\sigma_{\rm
  dc}=A\exp(-E_{\rm dc}/k_{\rm B}T)$, where $A$ is the pre-exponential
factor and $E_{\rm dc}$ the activation energy.  In disordered electron
conductors, variable range hopping can be relevant, leading to Mott's
law, $\ln\sigma_{\rm dc}\sim -1/T^{1/4}$.  The dielectric strength
$\Delta\epsilon$ depends weakly on temperature, often according to a
Curie law, $\Delta\epsilon\propto 1/T$.  Scaling functions $F(.)$ are
very similar for different materials, albeit not exactly the same.

A scaling behavior according to Eq.~\eqref{eq:sigma-BNN-scaling}, with
$\Delta\epsilon$ replaced by a related factor, was predicted
theoretically and verified for materials \cite{Summerfield:1985} based
on the extended pair approximation \cite{Summerfield/Butcher:1982} for
hopping transport and later interpreted by refined theoretical
approaches \cite{Hunt:1992a, Hunt:1992b, Baranovskii/Cordes:1999}.
Dyre and Schr{\o}der showed that Eq.~\eqref{eq:sigma-BNN-scaling} can
be accounted for by the Random Barrier Model (RBM) \cite{Dyre:1988,
  Dyre/Jacobsen:1995, Dyre/Schroeder:1996, Roling:2001b,
  Dyre/Schroeder:2002, Schroeder/Dyre:2002}, where independent
particles perform thermally activated hops over random, uncorrelated
energy barriers of a lattice.  For dimensions larger than two at low
temperatures, corresponding to an ``extreme disorder limit'', they
derived the relation \cite{Schroeder/Dyre:2008}
\begin{equation}
\ln\tilde\sigma=\frac{i\tilde\omega}{\tilde\sigma}
\left(1+2.66\,\frac{i\tilde\omega}{\tilde\sigma}\right)^{-1/3}
\label{eq:sigma-scaling-RBM}
\end{equation}
between the normalized conductivity $\tilde\sigma=\sigma/\sigma_{\rm
  dc}$ and scaled frequency
$\tilde\omega=\epsilon_0\Delta\epsilon\,\omega/\sigma_{\rm dc}$ based
on an improved variant of the diffusion-cluster-approximation
\cite{Schroder/Dyre:2000}.  Equation~\eqref{eq:sigma-scaling-RBM}
gives the scaling function $F(.)$ in the RBM very accurately and
provides a good approximation also for scaling functions of real
materials.

In amorphous solids, disorder is present also in energies of sites,
which represent potential wells where charge carriers reside for a
long time before passing a barrier in a rare thermally activated
transition.  This site energy disorder was shown to be important, in
particular, for understanding compositional effects in ion conducting
glasses, as, e.g., the mixed alkali \cite{Day:1976, Maass/etal:1992,
  Balasubramanian/Rao:1993, Habasaki/etal:1995, Maass:1999, Hunt:1999,
  Kirchheim:2000, Swenson/etal:2001, Ingram/Roling:2003,
  Swenson/Adams:2003, Habasaki/etal:2004, Lammert/Heuer:2005,
  Peibst/etal:2005, Maass/Peibst:2006, Kumari/etal:2012,
  Tsuchida/etal:2017, Shan/etal:2019, Wilkinson/etal:2019,
  Noritake/Naito:2023} and mixed glass former effect
\cite{Desphande/etal:1988, Jayasinghe/etal:1996, Pradel/etal:2003,
  Anantha/Hariharan:2005, Zielniok/etal:2007, Raskar/etal:2008,
  Haynes/etal:2009, Schuch/etal:2009,
  Schuch/etal:2011,Larink/etal:2012, Shaw/Ghosh:2014,
  Karlsson/etal:2015, Larink/etal:2015, Martin/etal:2015b,
  Martin/etal:2019}.  The site energy disorder requires to take into
account particle interactions, because independent particles would
accumulate at the sites with lowest energy at low $T$.  In view of
this, one may wonder why the conductivity
scaling~\eqref{eq:sigma-BNN-scaling} can be successfully described by
the RBM, where sites have equal energy and a hopping motion of
independent particles is considered.

%%%%%%%%%%%%%%%%%%%%%%%%%%%%%%%%%%%%%%%%%
\begin{figure*}[t!]
\includegraphics[width=\textwidth]{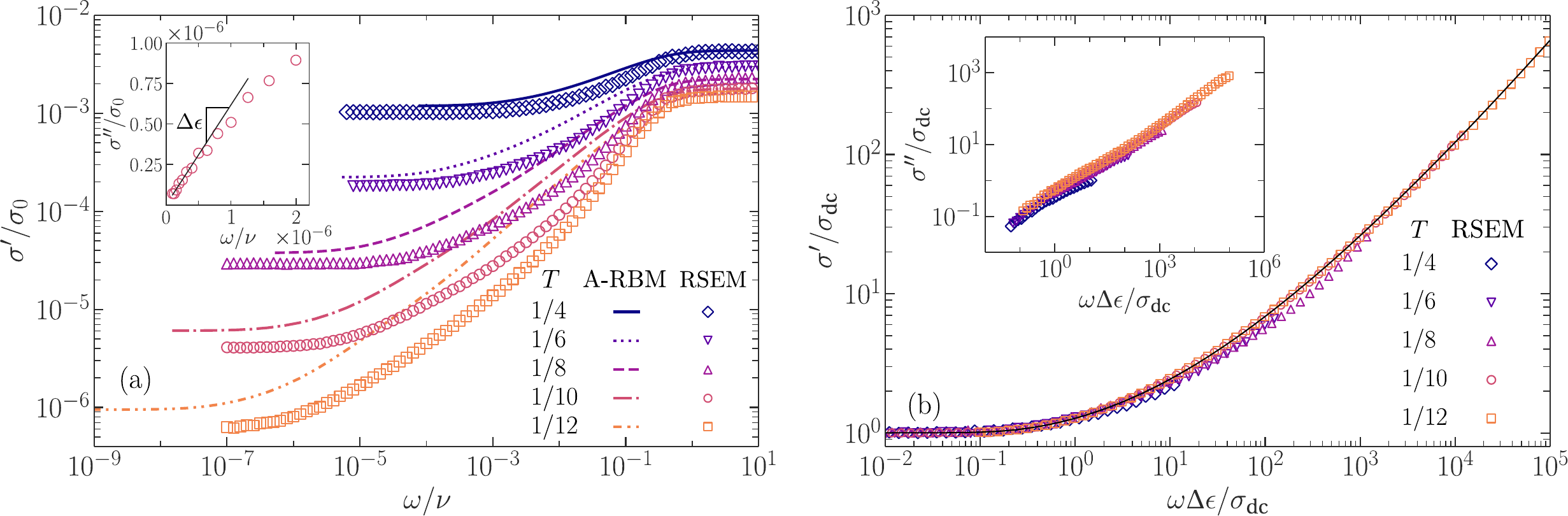}
\caption{(a) Real parts $\sigma'(\omega,T)$ of conductivity spectra
  for five different temperatures. Symbols refer to the RSEM and lines
  to the A-RBM. The inset shows the imaginary part
  $\sigma''(\omega,T)$ for $T=1/10$ on linear axes scales, where $\Delta\epsilon(T)$ is
  extracted from the low-frequency behavior $\sigma''(\omega,T)\sim
  \Delta\epsilon\,\omega$. (b) Scaled spectra for frequencies one
  order of magnitude below the attempt frequency, $\omega\le\nu/10$.
  The inset shows the correspondingly scaled imaginary parts of the
  conductivity. Data for the RSEM and A-RBM were obtained by averaging
  over 100 realizations of site energies. The solid line represents
  the master curve given by Eq.~\eqref{eq:sigma-scaling-RBM}.}
\label{fig:spectra_RSEM}
\end{figure*}
%%%%%%%%%%%%%%%%%%%%%%%%%%%%%%%%%%%%%%%%%

In this article we tackle this question by studying the Random Site
Energy Model (RSEM) with site-exclusion interaction, i.e.\ each
lattice site can be occupied by at most one particle. This model can
be viewed a minimal model for thermally activated particle transport
in disordered solids with site energy disorder. We are not considering variable-range 
hopping in this study, which would be important for electron tunneling.
With respect to
time-temperature scaling features, the RSEM was studied before in two
dimensions based on a linearized version of the master equation
\cite{Pasveer/etal:2006}, and in three dimensions by percolation
theory \cite{Hunt:1992b, Baranovskii/Cordes:1999} and simulations of
time-dependent diffusion coefficients \cite{Porto/etal:2000}.

The RSEM is fundamentally different from the RBM because it is a
many-particle model with site exclusion interaction.
This becomes immediately evident
when considering the variation of activation energies upon changing the concentration 
$n$ of mobile charge carriers in a fixed energy landscape: in the RBM with noninteracting particles, 
ion mobilities are independent of $n$. The
conductivity activation energy does not depend on ion concentration and the
conductivity is proportional to it. 
In contrast, the activation energy in the RSEM has a strong nonlinear dependence
on ion concentration.

This is just one fundamental difference between the RSEM and RBM. Another one concerns the preferential occupation of low energy
sites in the RSEM below the Fermi energy.
Even when comparing hopping dynamics of a single particle in a potential landscape with random traps
and saddle points on equal level, the two models are very different, because high energy barriers are mainly avoided in the diffusive motion of a single particle, while deep traps are not. In fact, the conductivity is independent of frequency in
random trap model of noninteracting particles \cite{Haus/Kehr:1987}.

Despite these fundamental differences, we here relate particle hopping in the RSEM to that
of an associated random barrier model (A-RBM) and demonstrate that
time-temperature scaling of conductivities in the A-RBM agrees with
that in the RSEM.  However, scaling in the RSEM sets in at scaled
temperatures $k_{\rm B}T/E_{\rm dc}$ higher than in the A-RBM, in
agreement with experimental observations. This suggests that site
energy variations and many-particle dynamics are important also to
understand time-temperature superposition in real materials.

\section{Scaling of conductivity spectra in the Random Site Energy Model}
\label{sec:scaling-RSEM}
For our analysis of the RSEM, we choose a simple cubic lattice with
lattice constant $a$ and assign to each site $i$ a random site energy
$E_i$ drawn from a Gaussian distribution with zero mean and standard
deviation $\Delta_E$. The particle concentration is $n=0.9\,a^{-3}$,
corresponding to a typical vacancy concentration in ion conducting
glasses \cite{Lammert/etal:2003, Habasaki/Hiwatari:2004,
  Mueller/etal:2007}. In the absence of an electric field, particles
jump from occupied sites $i$ to vacant nearest neighbor sites $j$ with
the Metropolis hopping rates
\begin{align}
w_{ij}&=\nu\min(1,\exp[-(E_j-E_i)/k_{\rm B} T])\nonumber\\[0.5ex]
&=w(\beta(E_i-E_j))\,,
\end{align}
where $\nu$ is an attempt frequency and $\beta=1/\kb T$.
We perform kinetic Monte Carlo (KMC) simulations
and determine 
$\sigma(\omega, T)$ by applying the method described in Ref.~\cite{Maass/etal:1995}.
Details are given in Appendix~\ref{app:KMC-simulations}.

As units of length, time and energy we use $a$, $\nu^{-1}$, and
$\Delta_E$ respectively, and as unit of conductivity $\sigma_0=nq^2\nu
a^2/\Delta_E$, where $q$ is the particle charge.

%%%%%%%%%%%%%%%%%%%%%%%%%%%%%%%%%%%%%%%
\begin{figure}[t!]
\includegraphics[width=\columnwidth]{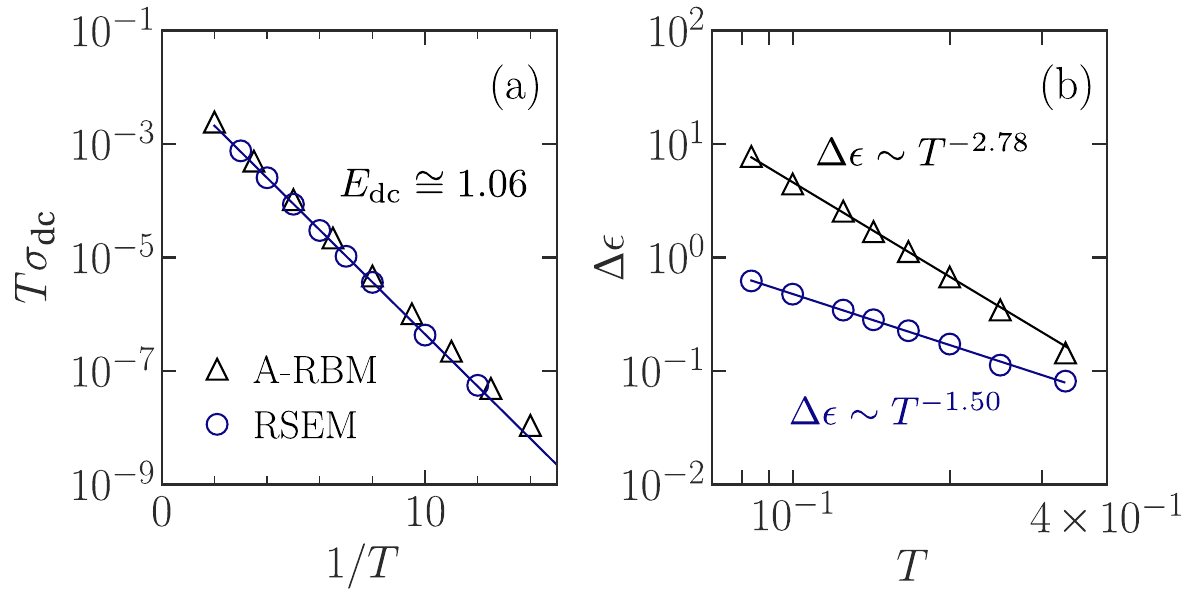}
\caption{(a) Arrhenius plots of dc-conductivities in the RSEM and
  A-RBM. A least squares fit to the data (solid line) yields an
  activation energy $E_{\rm dc}\cong1.06$. (b) Dielectric strength
  $\Delta\epsilon$ as a function of temperature for the two models.
  Least squares fits to power laws (solid lines, fit for $T\le1/6$ for
  the A-RBM) yield $\Delta\epsilon\sim T^{-1.50}$ (RSEM) and
  $\Delta\epsilon\sim T^{-2.78}$ (A-RBM).}
\label{fig:scaling_parameters}
\end{figure}
%%%%%%%%%%%%%%%%%%%%%%%%%%%%%%%%%%%%%%%

Figure~\ref{fig:spectra_RSEM}(a) shows $\sigma'(\omega,T)$ for the
RSEM (symbols) at five temperatures $T$.  These real parts of the
spectra monotonically increase with $\omega$ from the dc-value
$\sigma_{\rm dc}$ to the high-frequency plateau $\sigma_\infty$, where
the dispersion strength $\sigma_\infty/\sigma_{\rm dc}$ is the
stronger the lower the temperature.  From the linear increase
$\sigma''(\omega,T)\sim \Delta\epsilon\,\omega$ for $\omega\to0$, we
extract the dielectric strengths $\Delta\epsilon(T)$, see the inset of
Fig.~\ref{fig:spectra_RSEM}(a) for an example.

Figure~\ref{fig:spectra_RSEM}(b) shows that a scaling according to
Eq.~\eqref{eq:sigma-BNN-scaling} yields a data collapse onto a common
master curve for frequencies $\omega\le\nu/10$ one order of magnitude
smaller than the attempt frequency $\nu$. As $\nu$ is of the order of
phonon frequencies $\nu\simeq 10^{11-13}\,\si{s^{-1}}$, this regime
corresponds to that below $\sim100$~GHz, where scaling of conductivity
spectra is observed \footnote{Higher frequencies reflect short-time
  dynamics, where for modeling measured spectra one needs to take
  account the dynamics of the charge carriers in their potential
  wells, which is not covered by the hopping model.}.  We further
focus on this frequency range relevant for experiments.

The temperature dependence of the scaling parameters $\sigma_{\rm dc}$
and $\Delta\epsilon$ of the RSEM (circles) is displayed in
Fig.~\ref{fig:scaling_parameters}.  From a least squares fit to
$\sigma_{\rm dc}$ in the Arrhenius plot of
Fig.~\ref{fig:scaling_parameters}(a), we obtain an activation energy
$E_{\rm dc}\cong 1.06$.  Evaluating $T\sigma_{\rm dc}(T)\exp(E_{\rm
  dc}/k_{\rm B}T)$ for different $T$, we find the preexponential
$A\cong 0.018$ to be independent of $T$.  In the simulated temperature
range, the dielectric strength in the double-logarithmic plot of
Fig.~\ref{fig:scaling_parameters}(b) can be fitted by a power law
$\Delta\epsilon\sim T^{-1.50}$.

\section{Mapping of Random Site Energy onto Random Barrier Model}
For relating the RSEM to a barrier model, we use the theory of
Ambegaokar, Halperin and Langer (AHL theory)
\cite{Ambegaokar/etal:1971} According to this theory, dc-transport in
the RSEM is described by a mapping onto a random resistor network
\cite{Miller/Abrahams:1960} with symmetric link conductances.
\begin{equation}
g_{ij}=\beta q^2w_{ij} \langle n_i\rangle_{\rm eq}(1-\langle n_j\rangle_{\rm eq})=g_{ji}\,.
\label{eq:gij}
\end{equation}
Here, 
$\langle n_i\rangle_{\rm eq}$ are the mean occupation numbers in equilibrium,
\begin{equation}
\langle n_i\rangle_{\rm eq}=f(\beta(E_i-\mu))\,,
\end{equation}
with $f(x)=1/(e^x+1)$ the Fermi function and $\mu$ the chemical potential.

We extend the mapping to dispersive ac-transport by considering
independent particles to hop between nearest neighbor sites $i, j$
with rates $\nu \exp(-\beta U_{ij})$, where the energy barriers
$U_{ij}=U_{ji}$ are obtained by identifying the rate $w_{ij}\langle
n_i\rangle_{\rm eq}(1-\langle n_j\rangle_{\rm eq})$ in
Eq.~\eqref{eq:gij} with the hopping rate $\nu \exp(-\beta U_{ij})$:
\begin{align}
U_{ij}&=U(E_i,E_j)=
-\frac{1}{\beta}\ln\min(1,\exp[-\beta(E_j\!-\!E_i)])\nonumber\\
&\hspace{3em}{}-\frac{1}{\beta}\ln\bigl(f(\beta(E_i\!-\!\mu))[1\!-\!f(\beta(E_j\!-\!\mu))]\bigr)\,.
\label{eq:uij}
\end{align}
These barriers define the barrier landscape of the A-RBM. They depend
on temperature, and on the particle concentration $n$ via the chemical
potential $\mu$. 

By
setting the particle concentration equal to one in the A-RBM,  we ensure 
that $\sigma_\infty$ is equal to that in the RSEM. This is shown in Appendix~\ref{app:sigmainf}.

%\subsection{High-frequency conductivity}

%%%%%%%%%%%%%%%%%%%%%%%%%%%%%%%%%%%%%%%%%%%%%%%%%%%%%%%%%%
\begin{figure}[b!]
\includegraphics[width=\columnwidth]{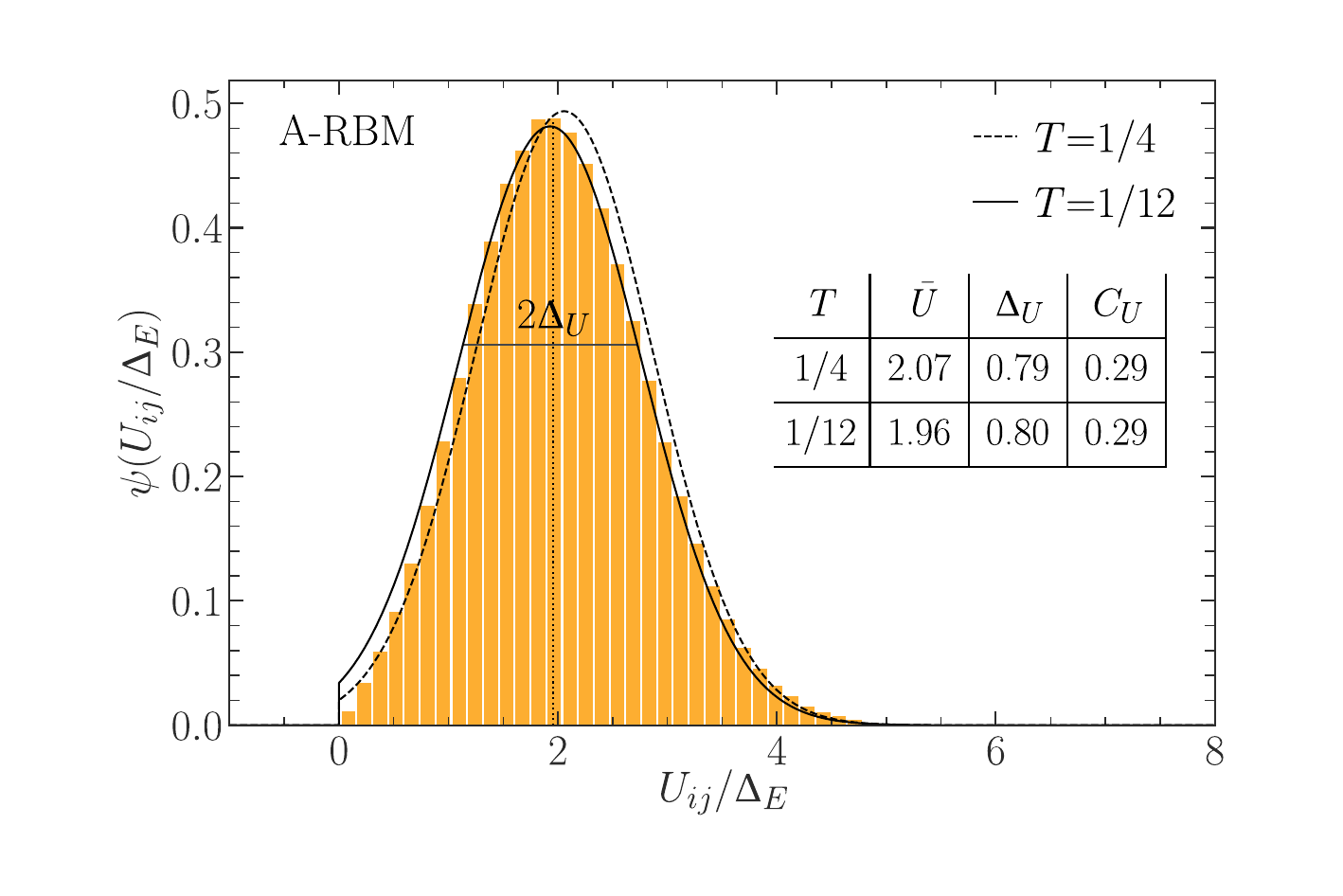}
\caption{Histogram of energy barriers $U_{ij}$ [Eq.~\eqref{eq:uij}] in
  the A-RBM for $T=1/12$. It can be described by a truncated Gaussian
  (solid line) with mean $\bar U$ and standard deviation $\Delta_U$ of
  the barrier distribution. Since $T=1/12$ is in the low-temperature
  regime (see text), the truncated Gaussian represents also the
  barrier distribution in the limit $T\to0$. At higher $T$, slight
  changes occur, as demonstrated by the truncated Gaussian describing
  the histogram at $T=1/4$.  The table gives $\bar U$, $\Delta_U$, and
  the correlation $C_U=\langle U_{ij}U_{kl}\rangle-\langle
  U_{ij}\rangle^2$ between barriers at neighboring links $(ij)$ and $(kl)$.}
\label{fig:ARBM_barrier_distributions}
\end{figure}
%%%%%%%%%%%%%%%%%%%%%%%%%%%%%%%%%%%%%%%%%%%%%%%%%%%%%%%%

\subsection{Barrier distribution and correlations}
Distributions of the barriers in the A-RBM are well described by
Gaussians truncated at $U=0$.  This is demonstrated for $T=1/4$ and
$T=1/12$ in Fig.~\ref{fig:ARBM_barrier_distributions}. The
distribution at the lower $T=1/12$ is representative for the $T\to0$
limit, where $U_{ij}=(|E_i-E_j|+|E_i-E_{\rm F}|+|E_j-E_{\rm F}|)/2$
and $E_{\rm F}=\mu(T\to0)=\sqrt{2}\Delta_E
\erf^{-1}(2n\!-\!1)\cong1.28$ is the Fermi energy.  That $T=1/12$
corresponds to the $T\to0$ limit can be understood by noting that for
a non-truncated Gaussian $\mu\sim E_{\rm F}[1+(\pi k_{\rm B}
  T/\sqrt{6}\Delta_E)^2]$ for $T\to0$, yielding a crossover
temperature $T_\times=\sqrt{6}\Delta_E/\pi k_{\rm B}\cong0.78$. Hence,
$T\ll T_\times$ for $T=1/12\cong0.083$.

The $U_{ij}$ of neighboring links are correlated even for spatially
uncorrelated site energies $E_i$. The autocorrelation $C_U=\langle
U_{ij}U_{kl}\rangle-\langle U_{ij}\rangle^2$ for neighboring links
$(ij)$ and $(kl)$ is given in the table in
Fig.~\ref{fig:ARBM_barrier_distributions}, together with the variance
$\Delta_U^2=\langle U_{ij}^2\rangle-\langle U_{ij}\rangle^2$ of the
barrier distributions.

\section{Comparison of conductivity scaling in A-RBM and RSEM}
\label{sec:spectra-ARBM}
For calculating $\sigma(\omega,T)$ 
of the A-RBM we apply the velocity autocorrelation (VAC) method \cite{Schroeder:2000}.
The determining equations are given in Appendix~\ref{app:VAC}. Real parts $\sigma'(\omega,T)$ are compared with the RSEM
in Fig.~\ref{fig:spectra_RSEM}(a), and data for $\sigma_{\rm dc}$ and $\Delta\epsilon(T)$ 
in Figs.~\ref{fig:scaling_parameters}(a) and (b).

In the temperature regime $1/12\le T\le 1/4$ of the data in
Fig.~\ref{fig:spectra_RSEM}, the dc-conductivity of the A-RBM
(triangles) agrees well with that of the RSEM (circles), see
Fig.~\ref{fig:scaling_parameters}(a).  This is surprising because
nearest-neighbor correlations between occupation numbers in the state
of constant current are factorized in the AHL theory. This
approximation, however, becomes worse at lower temperatures, causing
$\sigma_{\rm dc}$ of the A-RBM to differ from that of the RSEM.  The
increasing difference with lower $T$ is also visible when comparing
the low-frequency data in Fig.~\ref{fig:spectra_RSEM}(a).

%%%%%%%%%%%%%%%%%%%%%%%%%%%%%%%%%%%%%%%%%%%%%
\begin{figure}[b!]
\centering
\includegraphics[width=\columnwidth]{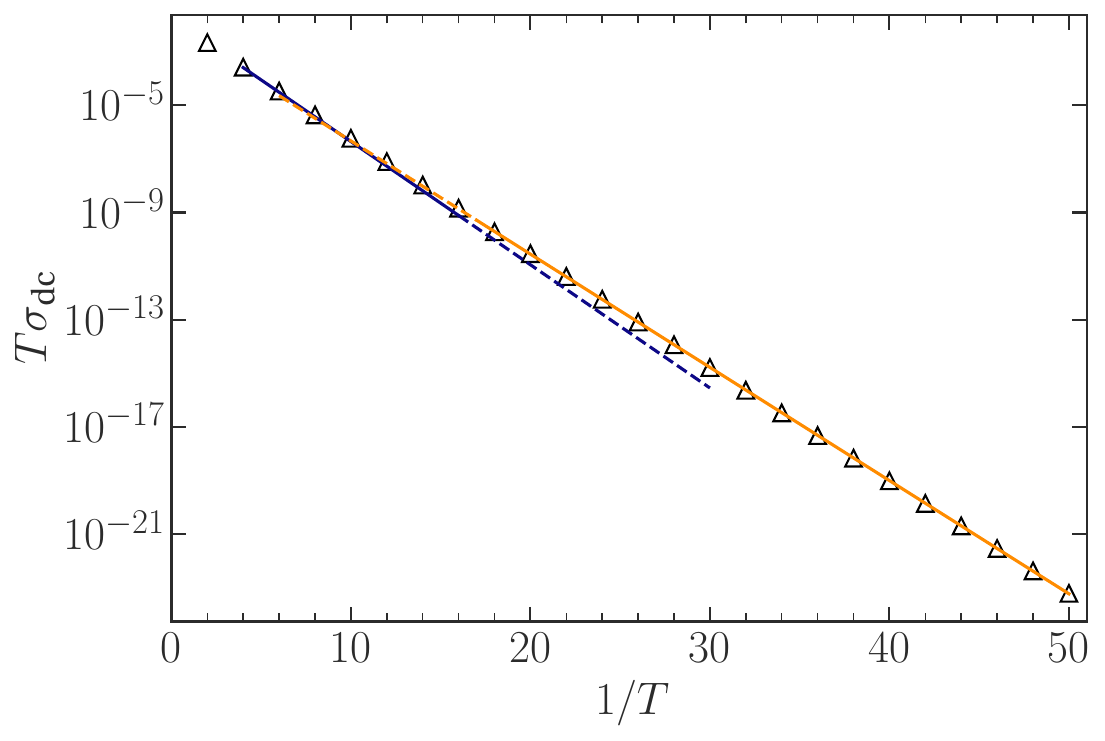}
\caption{\small Arrhenius plot of dc-conductivites of the A-RBM
  calculated from Eq.~\eqref{eq:sigma-VAC}. The activation energy
  $E_{\rm dc}\cong0.97$ obtained from a least squares fit in the range
  $16\le \beta\le 50$ (orange line) is in agreement with a percolation
  analysis based on the barriers in the limit $T\to0$
  [Eqs.~\eqref{eq:UijT=0},\eqref{eq:Edc-percanalysis}]. A least
  squares fit in the range $4\le \beta\le 15$ (blue line) yields the
  apparent activation energy $E_{\rm dc}\cong1.06$ that agrees with
  the activation energy of the RSEM in Fig.~2a.}
\label{fig:dc-cond-low-T}
\end{figure}
%%%%%%%%%%%%%%%%%%%%%%%%%%%%%%%%%%%%%%%%%%%%%

The VAC method allows one to calculate conductivities or the A-RBM
with high numerical accuracy at very low temperatures $T<10^{-2}$.
Arrhenius plots of dc-conductivities are shown in
Fig.~\ref{fig:dc-cond-low-T} for $\beta$ up to 50 and a least squares
fit in the range $16\le\beta\le50$ yields the activation energy
$E_{\rm dc}\cong 0.97$ (orange line). This value is in agreement with
a percolation analysis based on the zero-temperature limit of the
barriers in Eq.~\eqref{eq:uij},
\begin{equation}
U_{ij}(T\to0)=\frac{1}{2}\left(|\varepsilon_i-\varepsilon_j|
+|\varepsilon_i-\varepsilon_{\rms F}|+|\varepsilon_j-\varepsilon_{\rms F}|\right)\,.
\label{eq:UijT=0}
\end{equation}
According to this analysis, 
\begin{equation}
E_{\rm dc}={\rm min}\Bigl\{U_\times\,|\,\, 
\parbox[c]{0.57\columnwidth}{links $(ij)$ with $U_{ij}(T\to0)<U_\times$\\ form a percolating path}\Bigr\}\,.
\label{eq:Edc-percanalysis}
\end{equation}
We have determined the minimum 
by applying the Hoshen-Kopelman algorithm \cite{Hoshen/Kopelman:1976}
to large systems 
of $200^3$ sites, yielding a sharp percolation threshold at $\min\{U_\times\,|\,\ldots\}=E_{\rm dc}\cong0.97$.

This activation energy of the A-RBM is lower than $E_{\rm dc}\cong1.06$ 
in Fig.~\ref{fig:scaling_parameters}(a). Accordingly, $E_{\rm dc}$ 
in Fig.~\ref{fig:scaling_parameters}(a) is an apparent one for
the A-RBM. For the RSEM by contrast, our analysis suggests that
$E_{\rm dc}\cong 1.06$ is the true activation energy for $T\to0$.

The dielectric strength of the A-RBM has a temperature dependence
different from the RSEM. Data are presented in Fig.~\ref{fig:scaling_parameters}(b)
and can be fitted by a power law behavior $\Delta\epsilon\sim T^{-2.78}$ 
in the considered temperaturerange.

%%%%%%%%%%%%%%%%%%%%%%%%%%%%%%%%%%%%%%%%%%%%%%%%%%%%%%%
\begin{figure}[t!]
\includegraphics[width=\columnwidth]{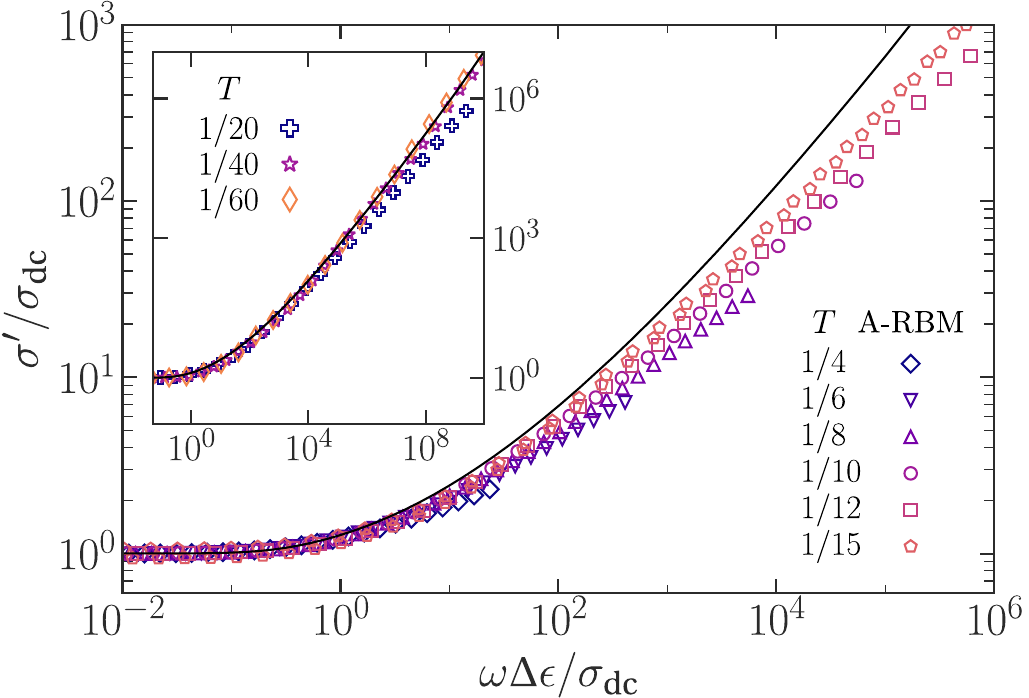}
\caption{Scaled real parts of conductivity spectra in the A-RBM for
  temperatures $T\ge1/15$, yielding no data collapse onto the master
  curve given by Eq.~\eqref{eq:sigma-scaling-RBM} (solid line).  The
  scaled data in the inset show that scaling behavior sets in at lower
  temperatures $T\lesssim1/40$.}
\label{fig:scaling_ARBM}
\end{figure}
%%%%%%%%%%%%%%%%%%%%%%%%%%%%%%%%%%%%%%%%%%%%%%%%%%%%%%%

Let us now consider the scaling behavior in the A-RBM.  Due to the
different temperature dependencies of $\Delta\epsilon$ and of
$\sigma_{\rm dc}$ at low $T$, the scaled frequencies
$\omega\Delta\epsilon/\sigma_{\rm dc}$ are different from that in the
RSEM.  Equation~\eqref{eq:sigma-scaling-RBM} was derived for
uncorrelated barriers, but short-range correlations between
neighboring barriers should not change the scaling form. This is
indeed true, as evident from the inset of Fig.~\ref{fig:scaling_ARBM},
where the data collapse at low $T\lesssim 1/40$.

However, the A-RBM shows no scaling for higher $T\gtrsim 1/40$, see in
particular the data in Fig.~\ref{fig:scaling_ARBM} for $1/12\le T\le
1/4$ corresponding to the temperature regime in
Fig.~\ref{fig:spectra_RSEM}.  Hence, scaled data for the A-RBM fall
onto the master curve at much lower $T$ than in the RSEM.

This finding is relevant when interpreting experiments.  For example,
scaling occurs for temperatures as high as $E_{\rm dc}/15k_{\rm B}$ in
sodium borate glasses \cite{Roling/etal:1997, Roling:1998} and $E_{\rm
  dc}/13k_{\rm B}$ in Bi$_2$O$_3$ doped P$_2$O$_5$–V$_2$O$_5$–MoO$_3$
electronically conducting nanocomposite glass \cite{Mandal/etal:2023},
corresponding to $T\simeq1/15$ and $T\simeq1/13$ in our dimensionless
units ($E_{\rm dc}\simeq 1$ in the RSEM and A-RBM). At such
temperatures, a good scaling is seen in Fig.~\ref{fig:spectra_RSEM}(b)
for the RSEM but not the A-RBM in Fig.~\ref{fig:scaling_ARBM}.

That scaling sets in at higher $T$ in the RSEM can be understood by
considering the high-frequency regime, where the scaled data of the
A-RBM in Fig.~\ref{fig:scaling_ARBM} lie significantly below the
master curve for $T\le 1/20$ in contrast to the corresponding data of
the RSEM in Fig.~\ref{eq:sigma-scaling-RBM}(b).  At high frequencies,
$\sigma(\omega)$ becomes increasingly dominated by forward-backward
hopping between sites with energies slightly above and below the Fermi
energy upon lowering $T$. The prevalence of this forward-backward
hopping at a given $T$ is more pronounced in the RSEM, where particles
tend to reoccupy a site below the Fermi level after leaving it.

To corroborate this interpretation, we have analyzed forward-backward
hopping probabilities $P_{\rm fb}$ along neighboring sites in both
models.  For determining $P_{\rm fb}$ in both the RSEM and A-RBM, 
we count the total number
$M_{\rm J}$ of jumps and the total number $M_{\rm fb}$ of
forward-backward jumps over a long simulation time starting from an
equilibrium state. The probability then is
\begin{equation}
P_{\rm fb}=\frac{M_{\rm fb}}{M_{\rm J}}\,.
\label{eq:Pfb}
\end{equation}
To obtain $M_{\rm fb}$, we store for each particle $i$ in the initial
equilibrium state the direction of its last jump. If particle $i$
performs a backward jump, a counter $m_i$ is incremented by one. If
the particle $i$ does not jump back, $M_{\rm fb}$ is incremented by
$m_i$, and $m_i$ reinitialized to zero.

%%%%%%%%%%%%%%%%%%%%%%%%%%%%%%%%%%%%%%%%%%%%%
\begin{figure}[t!]
\centering
\includegraphics[width=\columnwidth]{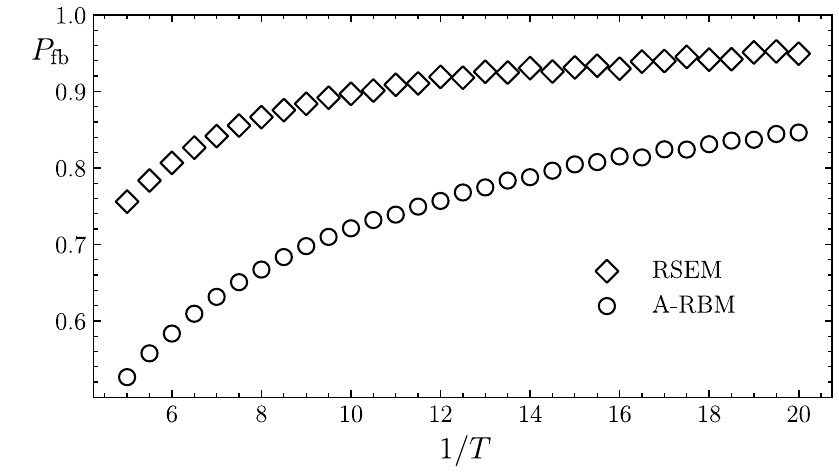}
\caption{\small Probability $P_{\rm fb}$ of forward-backward hopping [Eq.~\eqref{eq:Pfb}] 
in equilibrium as a function of $1/T$ for the RSEM and A-RBM.}
\label{fig:Pfb}
\end{figure}
%%%%%%%%%%%%%%%%%%%%%%%%%%%%%%%%%%%%%%%%%%%%%

Figure~\ref{fig:Pfb} shows $P_{\rm fb}$ as a function of $1/T$ for the
RSEM and A-RBM. The probability of forward-backward hopping is higher
for the RSEM and it increases with decreasing $T$ towards a saturated
value for both models. In the RSEM, saturation sets in for $T\lesssim
1/10$, indicating that the low-$T$-behavior is reached and scaled
conductivity data can be expected to fall on the master curve.  By
contrast, $P_{\rm fb}$ of the A-RBM increases much slower towards the
saturation limit upon lowering $T$. At $T=1/20$, it is still
significantly lower than in the RSEM.

\section{Conclusions}
\label{sec:conclusions}
In summary, we recovered the universal scaling behavior of
conductivity spectra for hopping motion between sites with randomly
varying energies. For the simplest case of site exclusion interaction,
we demonstrated how the many-particle dynamics can be mapped onto that
of independent particles hopping over barriers.  The associated
barrier model exhibits the same conductivity scaling as the original
site energy model. However, the scaling behavior is observed at higher
temperatures in the site energy model, which agrees with experimental
findings \cite{Roling/etal:1997, Roling:1998, Mandal/etal:2023}.

One may wonder why the scaling should not be affected when taking into
account the Coulomb interaction between the mobile charge carriers, or
peculiar features present in real materials such as complex geometries
of conduction paths. As for the Coulomb interaction, it is long-ranged
and can be viewed to give a weakly fluctuating contribution compared
to the variation of site energies, which is determined by the
interaction of the mobile charge carriers with the immobile host
network. Complex conduction paths arise naturally in the presence of
site or barrier energy disorder due to confinement of the transport at
low temperatures to percolating clusters with fractal geometry.

Our mapping of spatially fluctuating site onto barrier energies can be
extended to multicomponent systems. For mixed alkali glasses with two
types of mobile ions $\alpha={\rm A, B}$ having site energies
$E_i^\alpha$ and chemical potentials $\mu_\alpha$, barriers for
$\alpha$ ions are
\begin{align}
U_{ij}^\alpha=&\max(E_i^\alpha,E_j^\alpha)-\mu_\alpha\nonumber\\
&-\sum_{k=i,j}\frac{1}{\beta}
\ln\bigl[v(\beta(E_k^{\rm A}\!-\!\mu_{\rm A}),\beta(E_k^{\rm B}\!-\!\mu_{\rm B}))\bigr]\,,
\label{eq:uijalpha}
\end{align}
where $v(x,y)=1/(1+e^{-x}+e^{-y})$ for $x=\beta(E_i^{\rm A}\!-\!\mu_{\rm A})$
and $y=\beta(E_i^{\rm B}\!-\!\mu_{\rm B})$ is
the probability that site $i$ is vacant.

As a consequence, ions of type A and B experience distinct barrier
landscapes, leading to partial conductivities
$\sigma^\alpha(\omega,T)$ that scale with parameters
$\sigma^\alpha_{\rm dc}(T)$ and $\Delta\epsilon^\alpha(T)$ possessing
different temperature dependencies.  When the partial conductivities
are similar in magnitude, the total conductivity
$\sigma(\omega,T)=\sigma^{\rm A}(\omega,T)+\sigma^{\rm B}(\omega,T)$
is expected to show deviations from the universal time-temperature
scaling, as observed in experiments \cite{Cramer/etal:2002}. We
propose testing partial conductivity scaling in mixed ion
conductors. The generalized mapping \eqref{eq:uijalpha} furthermore
provides a new approach for a theoretical treatment of the mixed
alkali effect \cite{Leiber:2025}.

\begin{acknowledgments}
We gratefully acknowledge financial support by the Deutsche
Forschungsgemeinschaft (Project No.\ 428906592), and the use of a
high-performance computing cluster funded by the Deutsche
Forschungsgemeinschaft (Project No.\ 456666331).
\end{acknowledgments}

\appendix

\section{KMC simulations of RSEM and calculation of frequency-dependent conductivity}
\label{app:KMC-simulations}
For determining conductivity spectra $\sigma(\omega, T)$ of the RSEM,
we have performed KMC simulations by applying
the method described in Ref.~\cite{Maass/etal:1995}.  In this method,
the hopping of $N=nL^3\gg1$ particles in the presence of a sinusoidal
electric field $E(t)=E_0\sin(\omega t)$ in $x$-direction is considered
under periodic boundary conditions in all directions.  For small
amplitudes $E_0$ with $qE_0a\ll \kb T$, the effect of the field on the
hopping dynamics can be accounted for by choosing nearest-neighbor
target sites for jump attempts in $\pm x$-direction with
time-dependent probabilities.  In each elementary step, one of the $N$
particles is chosen randomly and attempted to move to a
nearest-neighbor site in $\pm x$-direction with probabilities $[1\pm
  qE(t)a/\kb T]/6$, and in $\pm y$- and $\pm z$-direction with
probability 1/6.  A jump attempt from site $i$ to a vacant nearest-neighbor
site $j$ is accepted with probability $\min(1,\exp[-(E_j-E_i)/k_{\rm
    B} T])$, and rejected if site $j$ is occupied.  
After each jump attempt, time is incremented by $\nu/N$.

Due to the applied electric field, a current in $x$-direction is
generated that becomes stationary after a transient time.  The current
density in the stationary state is calculated by subdividing a period
$2\pi/\omega$ of the electric field into $N_{\rm per}$ equidistant
time intervals $[t_k-\Delta t/2,t_k+\Delta t/2)$, $k=1,\ldots,N_{\rm
    per}$, where $t_k=k\Delta t-\Delta t/2$ and $\Delta
  t\ll2\pi/\omega$.  By counting the number $N_k^+$ and $N_k^-$ of
  particle jumps in the $+x$ and $-x$ direction in each interval $k$,
  we obtain the current densities
\begin{equation}
J_k=\frac{qa}{L^3}\,\frac{(N_k^+-N_k^-)}{\Delta t}
\end{equation}
at times $t_k$ in this period. Averaging over many periods in the
stationary state gives the period-averaged current densities $\langle
J_k\rangle_{\rms per}$, and a further averaging over many realizations
of the site disorder yields the period- and disorder-averaged current
densities ${\overline{\langle J_k\rangle}}_{\rms per}$. Real and
imaginary parts of the frequency-dependent conductivity are then
obtained from a linear least squares fit to
\begin{equation}
{\overline{\langle J_k\rangle}}_{\rms per}
=\sigma'(\omega)E_0\sin(\omega t_k)+\sigma''(\omega)E_0\cos(\omega t_k)\,.
\end{equation}

\section{Agreement of high-frequency conductivities in A-RBM and RSEM}
\label{app:sigmainf}
%By setting the particle concentration equal to one in the A-RBM, we
%ensure that $\sigma_\infty$ is equal to that in the RSEM. This can be
%seen as follows.
For hopping conduction in general, $\sigma_\infty$ is proportional to
the mean jump rate $\bar w_{\rm par}$ of a particle in the equilibrium
state \cite{Dieterich/etal:1980},
\begin{equation}
\sigma_\infty=\frac{\beta q^2a^2 n}{6}\,\bar w_{\rm par}\,.
\label{eq:siginf-barw}
\end{equation}
The rate $\bar w_{\rm par}^{\rms RSEM}$ in the equilibrium state of
the RSEM is related to the mean rate $\bar w_{\rms Link}^{\rms RSEM}$
for a link to be passed by a particle by
\begin{align}
\bar w_{\rm par}^{\rms RSEM}&=\frac{1}{N}\sum_i\sum_{j\,{\rms NN}\,i}
\langle w_{ij} n_i\tilde n_j\rangle_{\rm eq}
\label{eq:w-wlink-RSEM}\\
&\hspace{-2em}=\frac{6M}{N}\left[\frac{1}{6M}\sum_i\sum_{j\,{\rms NN}\,i} 
w_{ij} \langle n_i\rangle_{\rm eq} \langle\tilde n_j\rangle_{\rm eq}\right]
=\frac{6M}{N}\,\bar w_{\rms Link}^{\rms RSEM}\,,
\nonumber
\end{align}
where $M=(L/a)^3$ is the number of lattice sites.  and $\sum_{j\,{\rms
    NN}\,i}\ldots$ means summation over all nearest-neighbor sites $j$
of site $i$.  The mean link passage rate is
\begin{align}
\bar w_{\rms Link}^{\rms RSEM}&=\int\dd E\,\psi_E(E)f(\beta(E-\mu))
\label{eq:wlink-RSEM}\\
&\hspace{0.2em}{}\times\int\dd E'\psi_E(E')[1-f(\beta(E'-\mu))]\,
w(\beta(E'-E))
\nonumber
\end{align}
where $\psi_E(E)$ is the probability density of site energies $E$, chosen
to be the Gaussian $\psi_E(E)\propto\exp(-E^2/2\Delta_E^2)$ in our modeling.

In the A-RBM with $M$ noninteracting particles, the site occupation
probabilities are $p_i=M/M=1$ in the equilibrium state.  The mean jump
rate $\bar w^{\rms ARBM}_{\rms par}$ of a particle and the mean rate
$\bar w_{\rms Link}^{\rms ARBM}$ for a link to be passed by a particle
are related by
\begin{align}
\bar w^{\rms ARBM}_{\rms par}&
=\frac{1}{M}\sum_i\sum_{j\,{\rms NN}\,i} p_i \nu e^{-\beta U_{ij}}\nonumber\\
&=6\,\left[\frac{1}{6M}\sum_i\sum_{j\,{\rms NN}\,i} \nu e^{-\beta U_{ij}}\right]
=6\bar w_{\rms Link}^{\rms ARBM}\,.
\label{eq:w-wlink-ARBM}
\end{align}
The mean link passage rate is
\begin{equation}
\bar w_{\rms Link}^{\rms ARBM}=\int\hspace{-0.3em}\dd U\, \psi(U)\, \nu \exp(-\beta U)\,,
\label{eq:wlink-ARBM}
\end{equation}
where 
\begin{equation}
\psi(U)=
\int\hspace{-0.35em}\dd E\hspace{-0.25em}
\int\hspace{-0.35em}\dd E'\,\delta(U\!-\!U(E,E'))
\label{eq:psiU}
\end{equation}
is the probability density function of the assigned barriers.
Inserting Eq.~\eqref{eq:uij} into Eq.~\eqref{eq:psiU}, and
Eq.~\eqref{eq:psiU} into Eq.~\eqref{eq:wlink-ARBM}, yields
\begin{equation}
\bar w_{\rms Link}^{\rms ARBM}=\bar w_{\rms Link}^{\rms RSEM}\,.
\label{eq:wlink-equal}
\end{equation}
Using Eqs.~\eqref{eq:w-wlink-RSEM}, \eqref{eq:w-wlink-ARBM},
\eqref{eq:wlink-equal}, and the number densities $n=M/Ma^3=1/a^3$ and
$n=N/Ma^3$ for the A-RBM and RSEM, we obtain from
Eq.~\eqref{eq:siginf-barw}
\begin{align}
\sigma_\infty^{\rms ARBM}&=\frac{\beta q^2a^2}{6}\frac{1}{a^3}\,\bar w^{\rms ARBM}_{\rms par}
=\frac{\beta q^2}{a}\,\bar w_{\rm Link}^{\rms ARBM}
=\frac{\beta q^2}{a}\,\bar w_{\rm Link}^{\rms RSEM}\nonumber\\
&=\frac{\beta q^2}{a}\, \frac{N}{6M}\bar w_{\rm par}^{\rms RSEM}
=\frac{\beta q^2a^2 n}{6}\, \bar w_{\rm par}^{\rms RSEM}
=\sigma_\infty^{\rms RSEM}\,.
\label{eq:sigmainf-equal}
\end{align}
Hence the A-RBM with particle density $n=1/a^3$ has the same
high-frequency conductivity as the RSEM.

\section{Frequency-dependent conductivity of the A-RBM calculated from velocity autocorrelation method}
\label{app:VAC}
For calculating $\sigma(\omega,T)$ of the A-RBM we apply the velocity
autocorrelation (VAC) method.  This is a powerful numerical technique
to calculate ac conductivities in hopping transport of noninteracting
particles with symmetric jump rates \cite{Schroeder:2000}.  We here
give the relevant equations for determining $\sigma(\omega,T)$ of the
A-RBM.

For our statistically isotropic system, we take the field to be in
$x$-direction.  Conductivity spectra of the A-RBM with particle
number density $n=1/a^3$ are then given by
\begin{equation}
\sigma(\omega)=\frac{\beta q^2a^2}{M}\,i\omega\,\mathbf{1}_x^{\rm T}\,\mathbf{x}(\omega)\,,
\label{eq:sigma-VAC}
\end{equation}
where $\mathbf{1}_x$ and $\mathbf{x}(\omega)$ are column vectors with
$3M$ elements.  The first $M$ elements of the vector $\mathbf{1}_x$
are one and the other elements are zero.  The vector
$\mathbf{x}(\omega)$ is determined by the inhomogeneous system of
linear equations
\begin{equation}
(i\omega\mathbf{\Gamma}^{-1}\!\!+\!\mathbf{A}^{\hspace{-0.2em}\rm T}\mathbf{A})\,\mathbf{x}(\omega)=\mathbf{1}_x\,,
\label{eq:x-determ}
\end{equation}
where $\mathbf{\Gamma}$ is a $3M\times3M$ block diagonal matrix 
\begin{equation}
\mathbf{\Gamma}=\begin{pmatrix}
\mathbf{\Gamma}_1 & &\\
& \mathbf{\Gamma}_2  & \\
& & \mathbf{\Gamma}_3
\end{pmatrix}
\label{eq:Gammablock}
\end{equation}
consisting of $M\times M$ matrices $\mathbf{\Gamma}_k$ in the
diagonal, and $\mathbf{A}$ is a $M\times 3M$ block row matrix
$\mathbf{A}=(\mathbf{A}_1,\mathbf{A}_2,\mathbf{A}_3)$ with $M\times M$
matrices $\mathbf{A}_k$.  The $\mathbf{\Gamma}_k$ are the $M\times M$
diagonal matrices
\begin{equation}
\mathbf{\Gamma}_k=\begin{pmatrix}
\Gamma_{\bm{i}(1)}^{k+} & \cdots &  0\\
& \ddots & \\
0 & \cdots & \Gamma_{\bm{i}(M)}^{k+}
\end{pmatrix}\,.
\end{equation}
with $\Gamma_{\bm{i}(j)}^{k+}=\nu\exp[-\beta U_{\bm{i}(j)}^{k+}]$ the
jump rate from site ${\bm{i}(j)}$ in positive $k$ direction and
$U_{\bm{i}(j)}$ the barrier to be surmounted for a corresponding jump.
The multindex ${\bm i}(j)=(i_1(j),i_2(j),i_3(j))$ gives the position
of a site in the three-dimensional array and $j$ indices the same site
after a vectorization of the array with $i_1$ the fastest running
index, i.e.\
\begin{equation}
j=i_1+(i_2-1)M_0+(i_3-1)M_0^2\,,
\end{equation}
$i_k=1,\ldots,M_0$, where $M_0=L/a=M^{1/3}$ is the number of lattice
sites in along one coordinate axis.

The $\mathbf{A}_k$ are representing matrices for a discrete derivative
in $k$th direction under periodic boundary conditions.  Specifically,
\begin{equation}
\mathbf{A}_1=\mathbf{I}\otimes\mathbf{I}\otimes\mathbf{A}_0\,,\hspace{0.5em}
\mathbf{A}_2=\mathbf{I}\otimes\mathbf{A}_0\otimes\mathbf{I}\,,\hspace{0.5em}
\mathbf{A}_3=\mathbf{A}_0\otimes\mathbf{I}\otimes\mathbf{I}\,,
\end{equation}
where $\otimes$ denotes the Kronecker product operation, $\mathbf{I}$
is the $M_0\times M_0$ identity matrix, and $\mathbf{A}_0$ is the
$M_0\times M_0$ matrix
\begin{equation}
\mathbf{A}_0=\begin{pmatrix}
1 & 0 & 0 & 0 & \cdots &  0 & -1\\
-1 & 1 & 0 & 0 & \cdots &  0 &0\\
0 & -1 & 1 & 0 & \cdots &  0 &0\\
 &  &  &  & \vdots & &\\
0 & 0 & 0 & 0 & \cdots &  -1 & 1
\end{pmatrix}\,.
\end{equation}

Equation~\eqref{eq:x-determ} can be solved numerically with high computational efficiency 
by linear equation solvers even for large $\beta=1/\kb T$.

%\bibliography{/Users/Maass/bibfiles/ionicconductors}
%\bibliography{ionicconductors-local}

%Control: production of eprint (1) enabled
%

\end{document}